\documentstyle[multicol,prb,aps]{revtex}
\begin{document}
\input epsf
\title{Transport in multilevel quantum dots: from the Kondo effect to
the Coulomb blockade regime} 

\author{A. Levy Yeyati, F. Flores and A. Mart\'{\i}n-Rodero}

\address{Departamento de F\'\i sica Te\'orica de la Materia Condensada C-V, 
Universidad Aut\'onoma de Madrid, E-28049 Madrid, Spain}

\date{\today}

\maketitle

\begin{abstract}
A new theoretical method is introduced to study coherent
electron transport in an interacting multilevel quantum dot. The method
yields the correct behavior both in the limit of
weak and strong coupling to the leads, giving a unified description
of Coulomb blockade and the Kondo effect. Results for the density of states
and the temperature dependent conductance for a two-level dot
are presented. The relevance of these results in connection to recent
experiments on the Kondo effect in semiconducting 
quantum dots \cite{Goldhaber,Kowenhoven} is discussed. 
\end{abstract}


\begin{multicols}{2}

The proper description of coherent electron transport in the presence of
strong electron-electron interactions has been one of the central issues
in the field of mesoscopic systems \cite{review}.
Semiconducting quantum dots (QDs) provide an almost ideal system where
the predictions of the theory can be tested and new effects could be searched
for. For instance, recent experiments have demonstrated the possibility
of exploring the Kondo effect, a prototypical correlation effect, using
this technology \cite{Goldhaber,Kowenhoven}.

From the theoretical point of view, correlation effects 
have been mainly analyzed in this kind of systems by means of the simple
Anderson model
with a single spin-degenerate level \cite{Ng,us}. Within this model the
Kondo effect arises due to fluctuations in the spin of an unpaired
electron \cite{Matveev}.
So far, there has been few attempts to include a multilevel
spectrum for describing either single QDs with quasi-degenerate 
levels or coupled QDs beyond a rate equation approach 
\cite{multilevel}.
The actual situation in semiconducting QDs should require the
inclusion of a multilevel spectrum whenever many single-particle states
with a small level separation
are involved in the transport process. This could be the 
case in experiments measuring the phase of the transmission amplitude through
a QD in the Coulomb blockade (CB) regime \cite{Yacoby} and also
in recent experiments on the Kondo effect \cite{Goldhaber,Kowenhoven}.

The aim of this letter is to introduce a new theoretical approach for
describing correlation effects in multilevel QDs. The approach is
constructed to yield the correct behavior both in the limit
of infinite and vanishing charging energy. This is achieved by introducing
an interpolative self-energy for the one-electron Green functions, an
approach which has been successfully applied to several interacting
systems including the equilibrium and non-equilibrium Anderson models 
\cite{us,Alvaro}. This
type of approach has been recently rediscovered and applied to analyze 
the Mott transition in the Hubbard model for arbitrary band filling
\cite{Kotliar,Potthoff}.

For describing the multilevel QD we consider a model Hamiltonian which
is a generalization of the single-level Anderson model, $H = H_{\mbox{dot}} 
+ H_{\mbox{leads}} + H_T$, where $H_{\mbox{dot}} = \sum_{m} \epsilon_m
\hat{d}^{\dagger}_m \hat{d}_m + U \sum_{l > m} \hat{n}_m \hat{n}_l$ 
corresponds to the uncoupled QD ($\hat{n}_m = \hat{d}^{\dagger}_m \hat{d}_m$); 
$H_{\mbox{leads}} = \sum_{k \in L,R} \epsilon_k \hat{c}^{\dagger}_k \hat{c}_k$ 
to the uncoupled leads, and $H_T = \sum_{m, k \in L,R}
t_{m,k} \hat{d}^{\dagger}_m \hat{c}_{k} + h.c.$ describes the coupling
between the dot and the leads. The labels $m$ and $l$ in $H$ denote the 
different dot levels including spin quantum numbers. 

Our main objective is the determination of the dot retarded Green
functions from which the different level charges
and the linear conductance can be obtained. In a frequency representation
they can be written as
$G_m(\omega) = \left[ \omega - \epsilon^{HF}_m 
- \Sigma_m(\omega) - i \Gamma_{m,L} (\omega)  - i \Gamma_{m,R} (\omega) 
\right]^{-1}$, 
where $\epsilon^{HF}_m = \epsilon_m + U \sum_{l \ne m} n_l$ is the
Hartree-Fock level (we adopt the notation $n_l = <\hat{n}_l>$), 
$i \Gamma_{m,L(R)} (\omega) = \sum_{k \in L(R)}
t_{m,k}^2/(\omega - \epsilon_k + i0^+)$ are the hybridization
terms describing the 
coupling to the leads and $\Sigma_m(\omega)$ is a self-energy
that takes into account electron-electron
interactions beyond the Hartree-Fock approximation. We 
shall neglect the indirect coupling between dot states \cite{comment}
and adopt the usual approximation of $\Gamma_{m,L}$, $\Gamma_{m,R}$ 
being independent of energy. 
In our approach \cite{us,Alvaro}, we look
for an interpolative self-energy yielding the correct $U/\Gamma_m
\rightarrow 0, \infty$ limits. 

In the $U/\Gamma_m \rightarrow \infty$ or ``atomic" limit, $G_m(\omega)$
can be obtained using the equation of motion method \cite{eom}, which yields

\end{multicols}
\widetext
\begin{equation}
G^{(at)}_m(\omega) = \frac{<\prod_{l \ne m} (1 - \hat{n}_l)>}{\omega -
\epsilon_m + i0^+} + 
\sum_{l \ne m} \frac{<\hat{n}_l \prod_{(s \ne l) \ne m} (1 -
\hat{n}_s)>}{\omega - \epsilon_m - U + i0^+} + \;.\;.\;.\;
+ \frac{<\prod_{l \ne m} \hat{n}_l>}{\omega - \epsilon_m - (M-1)U + i0^+} ,
\end{equation}
\vspace{-5mm}
\begin{multicols}{2}
\noindent
where $M$ denotes the total number of one-electron levels in
$H_{\mbox{dot}}$.
In this expression all possible charge states of the dot give a
contribution. Their evaluation requires the knowledge of many particle
correlations $<\hat{n}_1 \hat{n}_2>$, $<\hat{n}_1 \hat{n}_2 \hat{n}_3>$, 
etc. However, for sufficiently large $U$ fluctuations 
in the dot charge with respect to the mean charge $\cal{N}$ by more than one 
electron become negligible and $G^{(at)}_m(\omega)$ is accurately
given by the three poles expression

\begin{eqnarray}
G^{(at)}_m(\omega)  \simeq 
\frac{A^m_{N-1}}{\omega - \epsilon_m - U(N-1) + i0^+} + \nonumber \\
\frac{A^m_N}{\omega - \epsilon_m  - U N + i0^+} +  
\frac{A^m_{N+1}}{\omega - \epsilon_m  - U(N+1) + i0^+}, 
\end{eqnarray} 

\noindent
where $N = Int[\cal{N}]$. In order to yield the exact first three
momenta of Eq. (1) the weight factors $A^m_N$ should satisfy 
the following sum rules

\begin{eqnarray}
A^m_{N-1} + A^m_{N} + A^m_{N+1}  =  1 \nonumber \\
(N-1) A^m_{N-1} + N A^m_{N} + (N+1) A^m_{N+1}  =  \sum_{l \ne m} n_l 
\nonumber \\
(N-1)^2 A^m_{N-1} + N^2 A^m_{N} + (N+1)^2 A^m_{N+1}  =  \nonumber \\
\sum_{l \ne m} n_l + <\hat{n} \hat{n}>_m, 
\end{eqnarray}

\noindent
where $<\hat{n} \hat{n}>_m = \sum_{(l \ne k )\ne m} <\hat{n}_l \hat{n}_k>$. 
The special case $N=0$ ($N=M-1$) has to be treated as $N=1$ ($N=M-2$). 
Notice that this expression for $G^{(at)}_m(\omega)$ is
fully determined by the average charges $n_l$ and the two-body
correlations $<\hat{n}_l \hat{n}_k>$. 
From $G^{(at)}_m(\omega)$ one can define an
``atomic'' self-energy, $\Sigma^{(at)}_m = \omega - 
\epsilon^{HF}_m - \left[G^{(at)}(\omega)\right]_m^{-1}$. Using Eqs.
(2) and (3) it can be shown that $\Sigma^{(at)}_m$ can be written as
the ratio of two polynomials in $\omega$ of the form 

\begin{equation}
\Sigma^{(at)}_m  = \frac{a_m U^2  (\omega - \epsilon_m + i0^+) + b_m U^3}
{(\omega - \epsilon_m + i0^+)^2 +  
c_m U (\omega - \epsilon_m + i0^+) + d_m U^2} ,
\end{equation} 

\noindent
where $a_m = ({\cal N} - n_m) \left[1 - ({\cal N} - n_m) \right] +
<\hat{n} \hat{n}>_m$; $c_m = {\cal N} - n_m - 3N$; $d_m = <\hat{n}
\hat{n}>_m + 3N^2 - 1 - (3N-1)({\cal N} - n_m)$ and
$b_m = N^2(1 - N) - ({\cal N} - n_m) d_m$.
On the other hand, in the $U/\Gamma_m \rightarrow 0$ limit
$\Sigma_m$ is given by the second order expression \cite{Alvaro}    

\begin{eqnarray}
\Sigma^{(2)}_m(\omega) & = & U^2  \sum_{l \ne m}
\int_{-\infty}^{\infty} d\epsilon_1 d\epsilon_2 d\epsilon_3 \; 
\frac{\tilde{\rho}_m(\epsilon_1) \tilde{\rho}_l(\epsilon_2) 
\tilde{\rho}_l(\epsilon_3)}{\omega + \epsilon_2 - \epsilon_1 - 
\epsilon_3 + i0^+} \nonumber \\
& & \left[ f_1 f_3 
\left(1 - f_2 \right) - \left(1 - f_1 \right) 
\left(1 - f_3 \right) f_2 \right], 
\end{eqnarray}

\noindent
where $f_i = f(\epsilon_i)$ denotes the Fermi distribution function 
at the leads,
and $\tilde{\rho}_m(\omega)$ are effective densities of states given by
$\pi \tilde{\rho}_m(\omega) = \Gamma_m /((\omega - \tilde{\epsilon}_m)^2 +
\Gamma_m^2)$. The effective levels $\tilde{\epsilon}_m$ are introduced in
order to fulfill the Fermi-liquid relations associated with charge
conservation (Friedel sum rule (FSR) \cite{Langreth}) as discussed
below. 

In order to determine an interpolative scheme between the two limits let
us first notice that $\Sigma^{(2)}_m \rightarrow U^2 \alpha_m/(\omega
- \tilde{\epsilon}_m)$, where $\alpha_m = \sum_{l \ne m}
\tilde{n}_l (1 - \tilde{n}_l)$, when $\Gamma_m/\omega \rightarrow 0$.
On the other hand, $a_m \rightarrow \alpha_m$ in the small $U$ limit and
thus $\Sigma_m^{(at)} \rightarrow U^2 \alpha_m /(\omega - \epsilon_m)$ in
this case.
These properties suggest to define the interpolative self-energy, 
replacing $\omega - \epsilon_m$ by $\alpha_m/\Sigma^{(2)}_m +
\Delta \epsilon_m$,
where $\Delta \epsilon_m = \tilde{\epsilon}_m - \epsilon_m$, 
in Eq. (4) for $\Sigma_m^{(at)}$, which yields 

\end{multicols}
\widetext
\begin{eqnarray}
\Sigma_m  = \left(\frac{\Sigma^{(2)}_m}{\alpha_m}\right)  
\frac{a_m + \left(a_m \Delta\epsilon_m/U + b_m/\alpha_m \right)
\left(\Sigma^{(2)}_m/U \right) } 
{ 1 + \left(2 \Delta\epsilon_m/U  + c_m \right) \left(\Sigma^{
(2)}_m/ \alpha_m U \right) + \left(c_m \Delta\epsilon_m/U + d_m \right)
\left(\Sigma^{(2)}_m/\alpha_m U\right)^2}. 
\end{eqnarray}
\vspace{-2mm}
\begin{multicols}{2}

This expression provides the generalization to the multilevel case of
the interpolative self-energy first introduced in \cite{Alvaro} for the
single level Anderson model. The single level case ($M=2$) is readily obtained
from Eq. (6) when $N=1$ taking $<\hat{n} \hat{n}>_m = 0$.
It is evident by construction that 
$\Sigma_m \rightarrow \Sigma^{(at)}_m$ when $U/\Gamma_m
\rightarrow \infty$ and
$\Sigma_m \rightarrow \Sigma^{(2)}_m$ in
the small $U$ limit.

The final step in this approach is to determine the level charges $n_m$,
the correlation functions $<\hat{n}_l \hat{n}_k>$ and the effective levels
$\tilde{\epsilon}_m$ selfconsistently. The charges and the correlations
functions are determined 
through the relations $n_m = -\int_{-\infty}^{\infty}
f(\omega) \mbox{Im}G_m(\omega) d\omega/\pi$ and

\begin{eqnarray}
\sum_{l \ne m} <\hat{n}_l \hat{n}_m> = \frac{-1}{\pi U}
\int_{-\infty}^{\infty} f(\omega)  
\mbox{Im} \left[ \Sigma_m(\omega) G_m(\omega) \right] d\omega .
\end{eqnarray}

Eq. (7) is an exact relation that follows from
the equation of motion of the retarded Green functions. 
It is important to stress that 
the self-consistent determination of the two-body correlations is
{\it essential} to get the correct values of the charge  
for large $U$. 
The effective level is determined by imposing the condition:

\begin{equation}
\int_{-\infty}^{\infty}  f(\omega) \mbox{Im} \left[ G_m(\omega) 
\frac{\partial \Sigma_m}{\partial \omega} \right] d \omega  = 0 , 
\end{equation}
\noindent
which at zero temperature reduces to the Luttinger theorem
\cite{Luttinger} ensuring the fulfillment of the FSR
\cite{Langreth} $n_m = -\mbox{Im}[\ln{G_m(E_F)}]/\pi$.  
In Ref. \cite{us} we showed that the condition of consistency between
the effective and the final charges is nearly equivalent to imposing the
fulfillment of the FSR for the simple Anderson model. Similar self-consistency
conditions have been proposed in Refs. \cite{Kotliar,Potthoff}. 

We have first applied this formalism to the 
case of a doubly degenerate level (four-fold degeneracy when 
including spin) which is a simple generalization of the single
level Anderson model. The inset in Fig. 1 shows the charge per level 
$n_m$ as a function of the leads Fermi energy $E_F$. 
For the case shown in Fig. 1, corresponding to $\epsilon_m = 0$,
$\Gamma_m/U = 0.075$
(same for all levels) and zero temperature, one can observe a modulation 
in the charge
resembling the typical Coulomb staircase of the $\Gamma_m/U \rightarrow 0$
limit (shown as dotted line for comparison). 

\begin{figure}[!th]
\begin{center}
\leavevmode
\epsfysize=5.5cm
\epsfbox{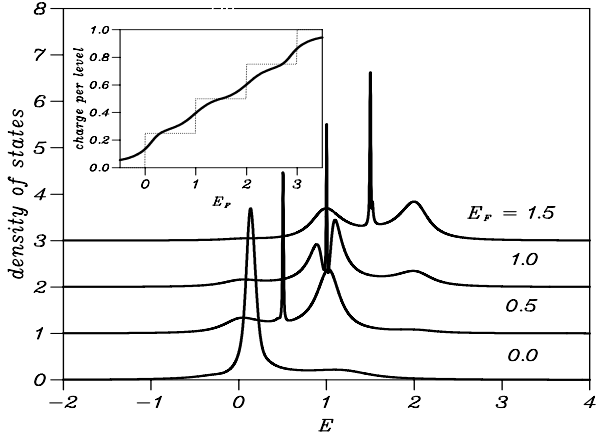}
\end{center}
\narrowtext
\caption[]{Density of states for a fully degenerate two-level dot with
$\Gamma_m/U = 0.075$ and $\epsilon_m = 0$ for different values of the leads 
Fermi energy
$E_F$. The inset shows the charge per level and spin as a function of $E_F$ 
(full line) as compared to the $\Gamma_m/U \rightarrow 0$ case (dotted line). 
Energies are measured in units of $U$.}
\label{fig1}
\end{figure}
 
Fig. 1 also shows the 
density of states (DOS) associated with the dot levels for different
values of $E_F/U = 0, 0.5, 1.0, 1.5$ corresponding to the steps and the
center of the plateaus in the charging curve. For $E_F/U=0$, $n_m
\simeq 0.125$ (the total charge on the dot being $\sim 0.5e$) the system
is in the so-called mixed-valence regime. The spectrum is in this case 
similar to
the one found using large $N$-expansions in the $U \rightarrow \infty$
limit \cite{Bickers} showing a resonance at a renormalized level 
just above $E_F$. 
For $E_F/U = 0.5$ and $1.5$ there are approximately one and two
electrons respectively inside the dot. For these cases the system is in
the Kondo regime, the DOS exhibiting a sharp peak around
$E_F$ (Kondo resonance) half way between two broader resonances 
(of width $\sim \Gamma$) approximately 
separated by $U$, corresponding to the elementary charge excitation
energy. 
It is worth noticing that the weight of these resonances is the same for
the $E_F/U=1.5$ case due to the electron-hole symmetry in the half-filled 
dot, while there is a pronounced asymmetry for $E_F/U=0.5$.         
For a half-integer occupation of the dot like in the $E_F/U=1.0$ case the
DOS exhibits a more complex structure with three broad
resonances around $E = 0, U$ and $2U$, and a narrow Kondo peak still
present at $E_F$. The overall shape is reminiscent of an average between the
cases with one and two electrons in the dot. 

The evolution of the Kondo peak as a function of $E_F$ and temperature 
should be reflected in the dot linear conductance, G, which can be readily
obtained from the Green functions (see Ref. \cite{Wingreen2}).
The conductance as a function of $E_F$ and different temperatures is
shown in Fig. 2. When approaching zero temperature the conductance
behaves like $\mbox{G} = 4 (e^2/h) \sin^2 \left[{\pi n_m}\right]$ 
as expected from the FSR. The conductance decreases very rapidly with
temperature in the region $0.5 < E_F < 2.5$ where the DOS at $E_F$ is
controlled by the Kondo peak. Outside this region one can notice a
slight increase of conductance with temperature.
At temperatures large enough to be above
the Kondo temperature, $T_K$, which can be estimated by the condition
$\mbox{G}(T_K) \simeq \mbox{G}(0)/2$ \cite{Goldhaber},
the conductance tends to exhibit the usual 
CB peaks at the charge degeneracy points. We should
remark that our approach does not provide an accurate estimate
of $T_K$ as a function of the model parameters as the exponential
decrease of the Kondo peak weight for very large $U$ \cite{Hewson} is
not strictly recovered. This limitation does not affect, however, the
qualitative behavior of the conductance except for a rescaling of the
temperature values.  

\begin{figure}[!th]
\begin{center}
\leavevmode
\epsfysize=5.5cm
\epsfbox{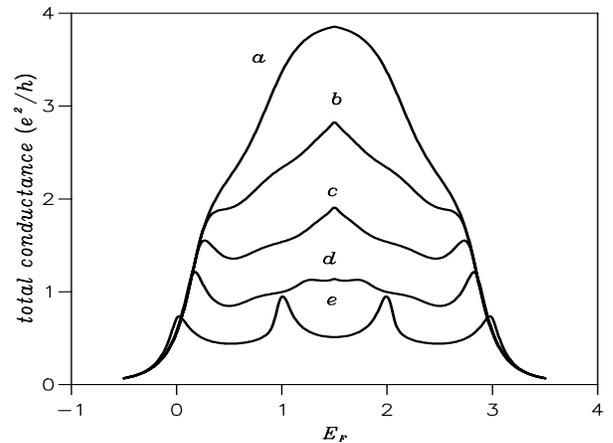}
\end{center}
\narrowtext
\caption[]{Total conductance for the same case of Fig. 1 as a function
of $E_F$ for different temperature values $T/U$: a) 0.0005, b)
0.0025, c) 0.005, d) 0.01 and e) 0.03.}
\label{fig2}
\end{figure}

So far we have analyzed the case of a completely degenerate level. 
In an actual QD geometrical asymmetries 
could result in an effective splitting $\Delta E$ of the dot levels. 
When $\Delta E \gg \Gamma$ the physical situation could be described by a
single level Anderson model. It is interesting to analyze the evolution
from this situation to the quasi-degenerate case $\Delta E \rightarrow
0$ previously discussed. This evolution is depicted in Fig. 3 and 4
where a splitting $\Delta E = 0.1 U$ and $\Delta E = 0.5 U$ is
introduced. Only the first half of the charging curve is shown as the
total conductance is symmetrical with respect to $E_F = (3U + \Delta E)/2$.
As can be observed in the charging curves shown as insets in these
figures, the splitting tends to block the charging of the upper levels.
This blocking effect is nearly complete for $\Delta E=0.5U$ and is
already apparent in the case of Fig. 3 where $\Delta E \sim \Gamma$.
The total conductance at low temperatures (case $a$) is observed to
decrease for increasing splitting approaching the 
behavior $2 (e^2/h) \sin^2[\pi n_m]$, where $m$ denotes the lower level.
The overall effect of temperature is again to reduce the total
conductance in the range for $E_F$ where the Kondo effect is present.
In the case of Fig. 4 the conductance {\it increases} with temperature in
the region $E_F \sim 1.5 U$ where the higher level starts to be
populated. One can also notice for this case that the conductance peaks
associated with the lower level are not completely symmetrical due to
the influence of the second level. 
Both effects, the increase of conductance with temperature between each
pair of peaks as well as the asymmetry, are present in the experimental
results of Ref. \cite{Goldhaber} and can be considered as a
manifestation of the multilevel structure of an actual QD.

\begin{figure}[!th]
\begin{center}
\leavevmode
\epsfysize=5.5cm
\epsfbox{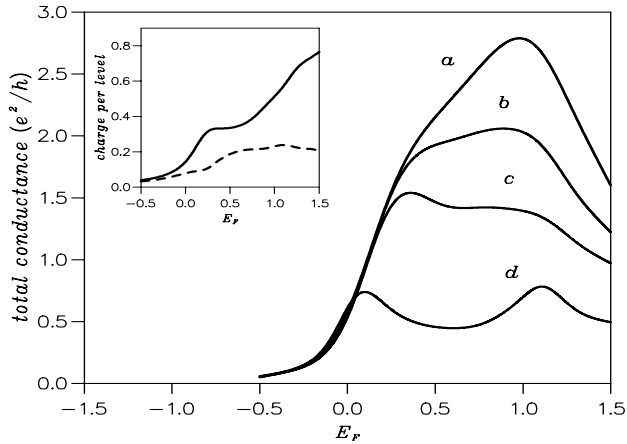}
\end{center}
\narrowtext
\caption[]{Total conductance for a two-level dot with splitting $\Delta
E/U = 0.1$ as a function of $E_F$ for different temperature values $T/U$: a)
0.0005, b) 0.0025, c) 0.005 and d) 0.03. The inset shows the charge per 
spin on the two levels as a function of $E_F$.
Only the first half of the curve is shown as the total conductance is
symmetrical with respect to $(3U + \Delta)/2$.}
\label{fig3}
\end{figure}

\begin{figure}[!th]
\begin{center}
\leavevmode
\epsfysize=5.5cm
\epsfbox{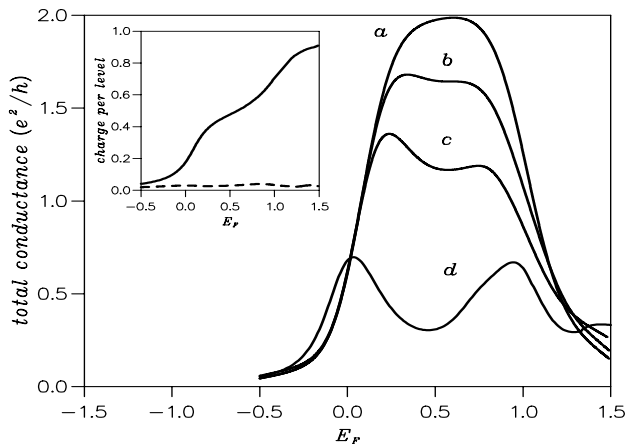}
\end{center}
\narrowtext
\caption[]{Same as Fig. 3 for  $\Delta
E/U = 0.5$.}
\label{fig4}
\end{figure}

The most remarkable consequences of this multilevel structure would be
observed in a situation corresponding to a smaller splitting 
where ``collective'' Kondo like features like the ones depicted in Figs.
2 and 3 should appear. Vertical dots with cylindrical symmetry, as 
those studied in Ref.  \cite{Tarucha}, constitute an almost ideal 
realization of the two-fold degenerate case. One word of caution should
be said, however, regarding the effects of Hund's rule, not included in
the present approach and which might introduce some deviation with respect
to the behavior depicted in Fig. 2. The results presented in this letter
will fully apply to a situation where the exchange 
interaction between the degenerate levels is much smaller than the Coulomb 
interaction $U$.
  
A.L.Y. and A.M.R. thank J.J. Palacios and C. Tejedor 
for fruitful discussions. This work has been funded by the Spanish CICyT under
contracts PB97-0028 and PB97-0044.

\end{multicols}

\end{document}